\documentclass[12pt,twoside,letter]{article}
\usepackage[utf8]{inputenc}
\usepackage[margin=2.5cm]{geometry}
\usepackage{amsmath,amsthm,mathtools,amssymb,lscape,mathabx}
\usepackage{thmtools,thm-restate}
\usepackage{mathrsfs}
\usepackage{proba}
\usepackage{booktabs}
\usepackage{longtable}
\usepackage{threeparttable,subfigure}
\usepackage{enumerate}
\usepackage[shortlabels]{enumitem}
\usepackage[round]{natbib}
\usepackage{graphicx,setspace,multirow}
\usepackage{dsfont}
\usepackage{caption}
\usepackage{xcolor}
\usepackage{chngcntr}
\usepackage{apptools}
\AtAppendix{\counterwithin{lemma}{section}}
\usepackage[english]{babel}
\usepackage{xr}
\usepackage{ragged2e,qtree,pict2e}
\usepackage{booktabs,caption}
\usepackage{threeparttable}
\usepackage{booktabs}
\usepackage{appendix}
\usepackage{chngcntr}

\usepackage{hyperref}
\hypersetup{
    unicode=false,          
    pdftoolbar=true,        
    pdfmenubar=true,        
    pdffitwindow=false,     
    pdftitle={},    
    pdfauthor={},     
    pdfsubject={},   
    pdfcreator={},   
    pdfproducer={},  
    pdfkeywords={}, 
    pdfnewwindow=true,      
    colorlinks=false,       
    linkcolor=red,          
    citecolor=green,        
    filecolor=magenta,      
    urlcolor=cyan           
}

\newtheorem*{theorem*}{Theorem}

\newtheorem*{example*}{Example}

\newtheorem{algorithm}{Algorithm}

\numberwithin{equation}{section}

\usepackage{siunitx}
\usepackage{xcolor}
\newcolumntype{R}{S[color=blue,negative-color = red]}

\def\b{\boldsymbol}

\def\1{\mathds{1}}

\definecolor{light-gray}{gray}{0.95}
\newcommand{\code}[1]{\colorbox{light-gray}{\texttt{#1}}}

\DeclareSymbolFont{sfoperators}{OT1}{cmss}{m}{n}
\DeclareSymbolFontAlphabet{\mathsf}{sfoperators}

\makeatletter
\def\operator@font{\mathgroup\symsfoperators}
\makeatother

\begin{document}

\title{Modeling and Forecasting Intraday Market Returns: a Machine Learning Approach\\
\vspace{0.2cm}}

\author{{\textbf{Iúri H. Ferreira}}\\
\small{Department of Economics}\\
\small{Pontifical Catholic University of Rio de Janeiro}\\ \small{E-mail:\texttt{
\href{mailto:iurihondaf@gmail.com}{iurihondaf@gmail.com}} 
} \and {\textbf{Marcelo C. Medeiros}} \\ 
\small{Department of Economics}\\
\small{Pontifical Catholic University of Rio de Janeiro}\\ \small{E-mail:\texttt{
\href{mailto:mcm@econ.puc-rio.br}{mcm@econ.puc-rio.br}} 
}}

\vspace{5pt}


\date{\small{\today}}

\maketitle

\begin{abstract}
\noindent 
In this paper we examine the relation between market returns and volatility measures through machine learning methods in a high-frequency environment. We implement a minute-by-minute rolling window intraday estimation method using two nonlinear models: Long-Short-Term Memory (LSTM) neural networks and Random Forests (RF). Our estimations show that the CBOE Volatility Index (VIX) is the strongest candidate predictor for intraday market returns in our analysis, specially when implemented through the LSTM model. This model also improves significantly the performance of the lagged market return as predictive variable. Finally, intraday RF estimation outputs indicate that there is no performance improvement with this method, and it may even worsen the results in some cases. 
\\

\vspace{5pt}

\vspace{5pt}

\vspace{5pt}
\noindent
\textbf{Acknowledgments}: This work is partially funded by CNPq and CAPES. 

\end{abstract}

\doublespacing

\newpage

\section{Introduction} \label{sec:introduction}




The recent advances in high-frequency data estimation are associated not only to the technological development and growing processing capacity of \textit{big data}, but also to the interest in understanding and predicting the behavior of variables in shorter time spaces. Machine Learning methods have been developed in parallel as increasingly accurate tools for estimating and predicting high-dimensional data. Both these fields of study can be easily accommodated into the economic and financial data environment. The adaptation of these methods to incorporate and update information over time allowed the development of robust predictive methods, progressively more relevant in time series analysis.

This paper examines the intrinsic relation between market returns and volatility measures, besides lagged returns themselves. Using minute-by-minute intraday data, we find that the CBOE Volatility Index (VIX) can be a strong predictor for the S\&P 500 ETF (SPY) in high-frequency, especially through machine learning models. For each day in the sample, we implement a 30 minute rolling-window estimation procedure to forecast the subsequent minute, totaling 340 estimated market returns in a day. This framework is based on the estimation scheme of \cite{Chinco2019} for a cross-section of stock returns. We use different approaches to estimate these returns, from ordinary least squares (OLS) regression benchmarks to more sophisticated methods, focusing on nonlinear models. To build minute-by-minute machine learning models, we relied on the work of \cite{MMM2021}.

Particularly, we apply Long-Short-Term Memory (LSTM) neural networks \cite{sHjS1997} and Random  Forest (RF) models \citet{breiman2001random} to  estimate  multiple intraday rolling windows, using a range of predictive variables in different settings. LSTM models are a variant of Recurrent Neural Networks (RNNs), that differ from the standard neural networks on the ability to remember the previous states in time. This method is broadly used in applications related to weather data, and speech and writing recognition. For economic and financial data in particular, we want this time dependence to be present, what makes the LSTM such an attractive model, specially due to its ability to remember what matters and forget what is irrelevant to the model. For a nice application of LSTMs in asset pricing see \citet{pelger19}. 

Introduced by \cite{breiman2001random}, the Random Forest is an ensemble method, which means it combines several simpler models, producing an optimal improved aggregate version in the end. In this case, the base models are classification or regression trees, a nonparametric method based on the mechanism of recursive partitioning of the space of covariates. The average of the estimates generated by each tree are used to build the final forecast, known as Random Forest. RF models have been shown to be a very competitive forecasting tool. See, for example, \citet{mcMgVaVeZ2019}.

The choice to analyze volatility measures as potential predictors is an important issue in the economic literature. Among others, \cite{Corsi2009} and \cite{PattonSheppard2015} established a whole literature on realized volatility ($RV$), defined as the daily sum of the cumulative squared returns during business hours of a trading day. The forecasting methodology in this field of research uses predominantly autoregressive structures endowed with long memory to predict realized volatilities. \cite{McAleerMedeiros2008} provide a extensive review of theoretical developments and empirical applications concerning realized volatility. 

Time-series properties of the VIX index are addressed by works like \cite{FMS2014}, as well as its positive contemporaneous link with the volume of the S\&P 500 index. \cite{Martin2017} and \cite{MartinWagner2019} explore the predictability of market and stock returns defining an associated volatility index named SVIX that provides a bound on the equity premium perceived by the investor. By associating implied variance (VIX$^2$) and realized variances measures, \cite{BTZ2009} introduce the \textit{variance risk premium} (VRP) as the difference between those two variables, that can be used to explain the aggregate stock market returns. \cite{HoerovaBekaert2014} extend these results by decomposing VIX$^2$ into a conditional variance of the stock market and the equity variance premium. The authors conclude that the variance premium is a significant predictor of stock returns.

This plethora of applications for volatility measures and machine learning methods raises our interest in combining these fields of study in a high frequency environment to predict future market returns. As will be discussed throughout the paper, our outputs indicate that machine learning models as the LSTM may improve the performance of benchmark linear regression models in a minute-by-minute framework. Furthermore, if we choose the right regressors, particularly the VIX in our study, the predictive ability becomes even higher. On the other hand, Random Forests did not bring improvements to our estimations, when compared to the established benchmarks.

The paper is organized as follows. Section \ref{sec:methodology} reviews the methodology to be implemented in the machine learning framework. The high-frequency data details are presented in Section \ref{sec:data}. Section \ref{sec:analysis} discusses the employment of different predictive models, as well as the benchmarks to our exercise. The results of intraday estimations and aggregate outputs are described in Section \ref{sec:results}. Finally, we conclude the paper in Section \ref{sec:conclusion}.

\section{Methodology} \label{sec:methodology}

Define the following high-frequency forecasting model, where $t$ is the day, and $m$ the intraday minute:
\begin{equation}\label{eq:model}
    Y_{t,m+h} = \mathcal{F}_h(\b{X}_{t,m}) + U_{t,m+h}, \quad h=1,\dots,H, \quad t=1,\dots,T
\end{equation}
where $\b{X}_{t,m} := (Y_{t,m-1},\dots,Y_{t,m-p},\b{Z}'_{t,m},\dots,\b{Z}'_{t,m-r})'$ is a $n$-dimensional vector of predictors, with $p \geq 1$ and $r \geq 0$. $\mathcal{F}_h: \mathbb{R}^n \rightarrow \mathbb{R}$ is an unknown measurable function and $U_{t,m+h} := Y_{t,m+h} - \mathcal{F}_h(\b{X}_{t,m})$ has zero mean and finite variance. 

For any chosen model framework and forecasting horizon $h=1,\dots,H$, we want to define the target function $\mathcal{F}_h$, to be estimated from the available data set. This function can be an ensemble of multiple models, and it can vary according to the horizon $h$.

\subsection{Nonlinear Models} \label{subsec:nonlinearmodels}

\cite{MMM2021} expatiate in detail about the different machine learning methods, as well as their advances and applications in time series data environment. The authors discuss both linear and nonlinear models, of which we will focus on the last. The reason for this is that sometimes the linearity hypothesis may not encompass all the characteristics of variables such as volatility measures, specially in a high-frequency environment. In this case, we look for alternatives in the universe of models in statistical learning literature.

\subsection{Long-Short-Term Memory Neural Networks}

Recurrent Neural Networks (RNNs) are neural networks that allow for feedback among the hidden layers. RNNs can use their internal state (memory) to process sequences of inputs. A generic RNN can be written as
\[
\begin{split}
\b{H}_{t,m} &= \b f(\b{H}_{t,m-1},\b{X}_{t,m}),\\
\widehat{Y}_{t,m+h|m} &= g(\b{H}_{t,m}),
\end{split}
\]
where $\widehat{Y}_{t,m+h|m}$ is the prediction of $Y_{t,m+h}$ given observations only up to minute $m$ at day $t$,  $\b f$ and $g$ are functions to be defined and $\b H_{t,m}$ is what we call the (hidden) state. From a time-series perspective, RNNs can be see as a kind of nonlinear state-space model.

\begin{figure}
    \centering
    \includegraphics[width=0.4\linewidth]{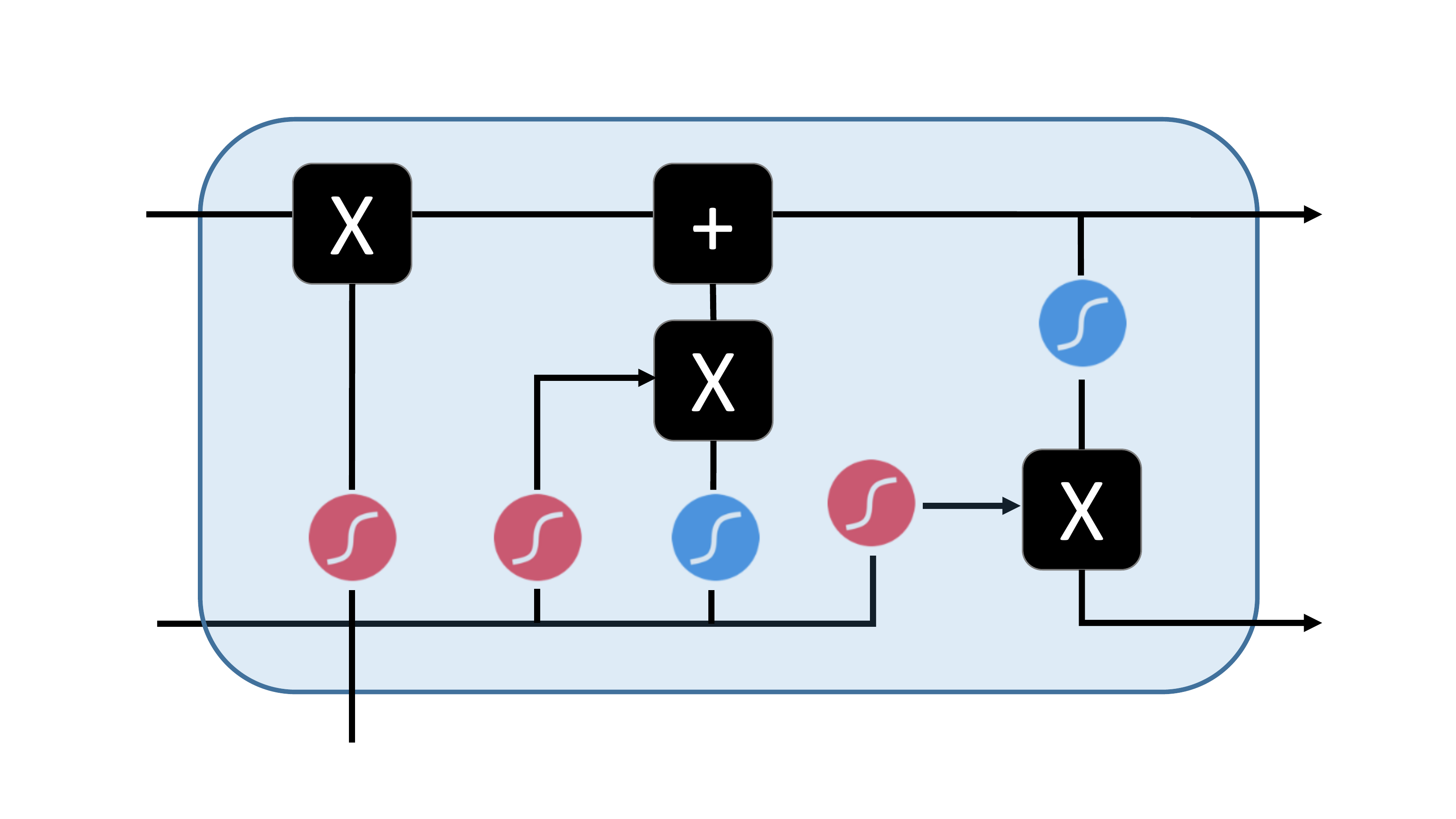}
    \includegraphics[width=0.4\linewidth]{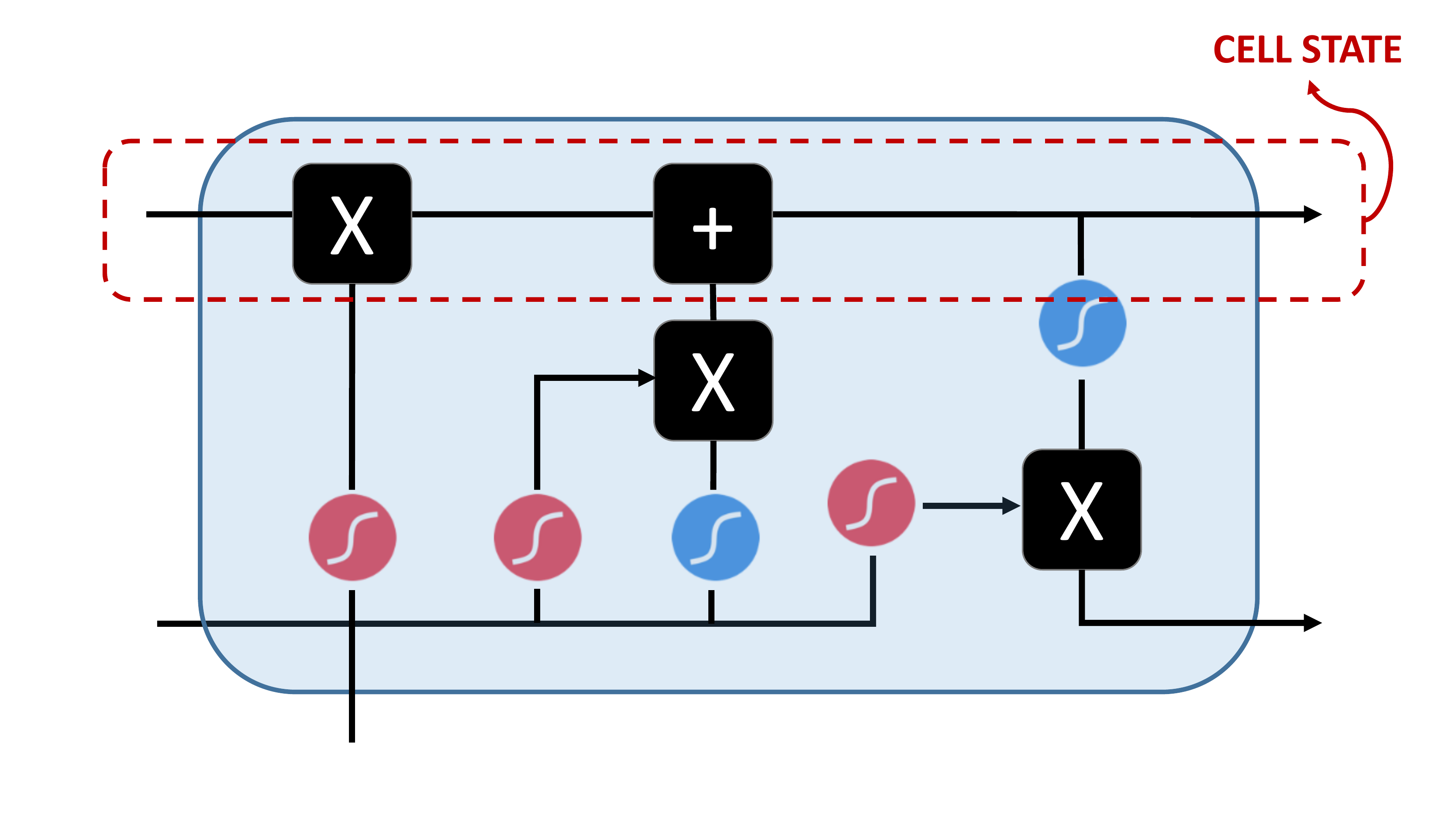}\\
    \includegraphics[width=0.4\linewidth]{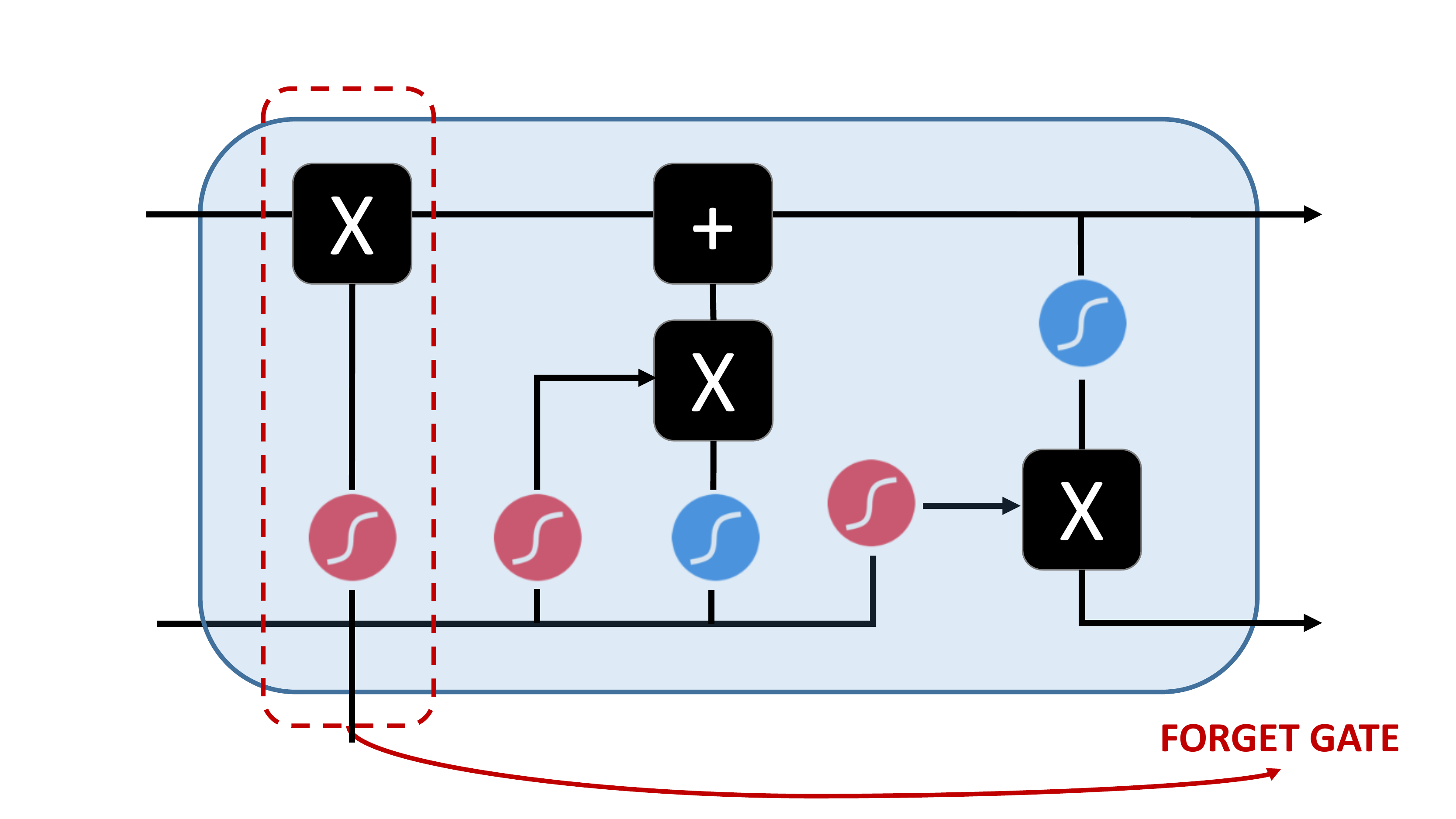}
    \includegraphics[width=0.4\linewidth]{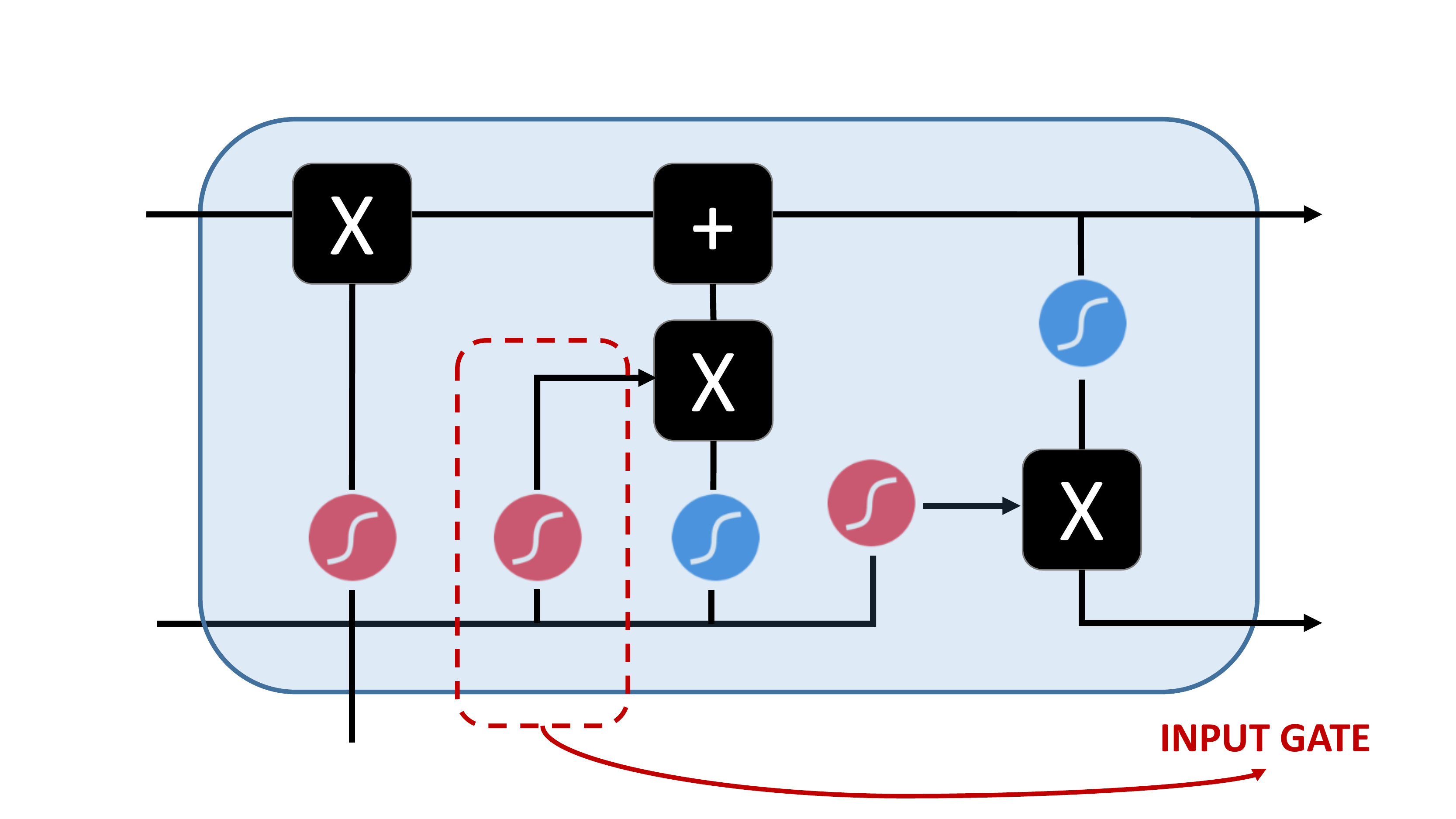}\\
    \includegraphics[width=0.4\linewidth]{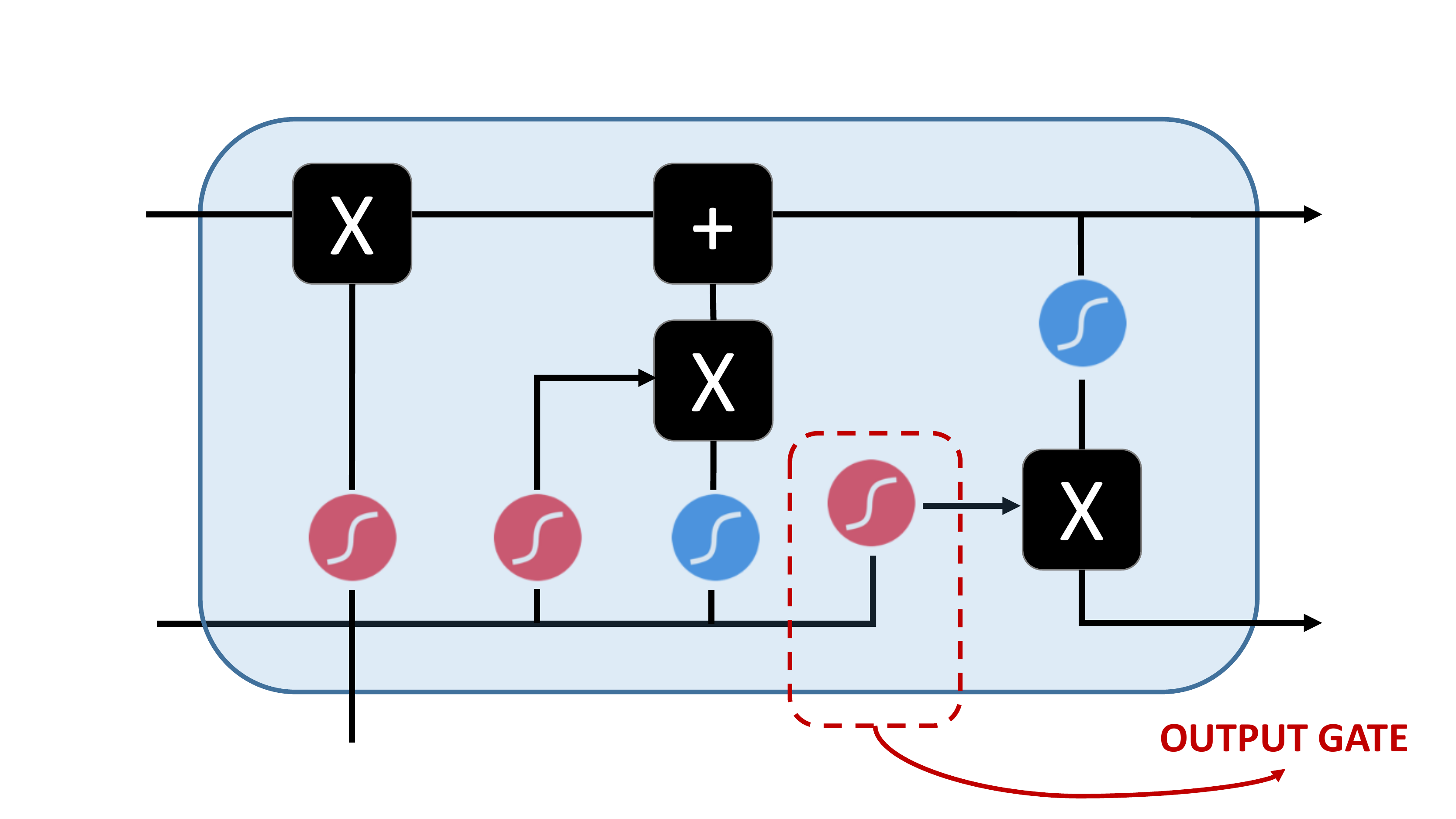}
    \caption{Architecture of the Long-Short-Term Memory Cell (LSTM)}
    \label{F:LSTM}
\end{figure}

RNNs can remember the order that the inputs appear through its hidden state (memory) and they can also model sequences of data so that each sample can be assumed to be dependent on previous ones. However, RNNs are hard to be estimated as they suffer from the vanishing/exploding gradient problem. Fortunately, there is a solution to the problem proposed by \citet{sHjS1997}: the Long-Short-Term Memory (LSTM) network . Figure \ref{F:LSTM} shows the architecture of a typical LSTM layer. A LSTM network can be composed of several layers. In the figure, red circles indicate logistic activation functions, while blue circles represent hyperbolic tangent activation. The symbols ``\textsf{X}'' and ``\textsf{+}'' represent, respectively, the element-wise multiplication and sum operations. The RNN layer is composed of several blocks: the cell state and the forget, input, and ouput gates. The cell state introduces a bit of memory to the LSTM so it can ``remember'' the past. LSTM learns to keep only relevant information to make predictions, and forget non relevant data. The forget gate tells which information to throw away from the cell state. The output gate provides the activation to the final output of the LSTM block at day $t$ and minute $m$. Usually, the dimension of the hidden state ($\b{H}_{t,m}$) is associated with the number of hidden neurons.

\begin{figure}
    \centering
    \includegraphics[width=0.9\linewidth]{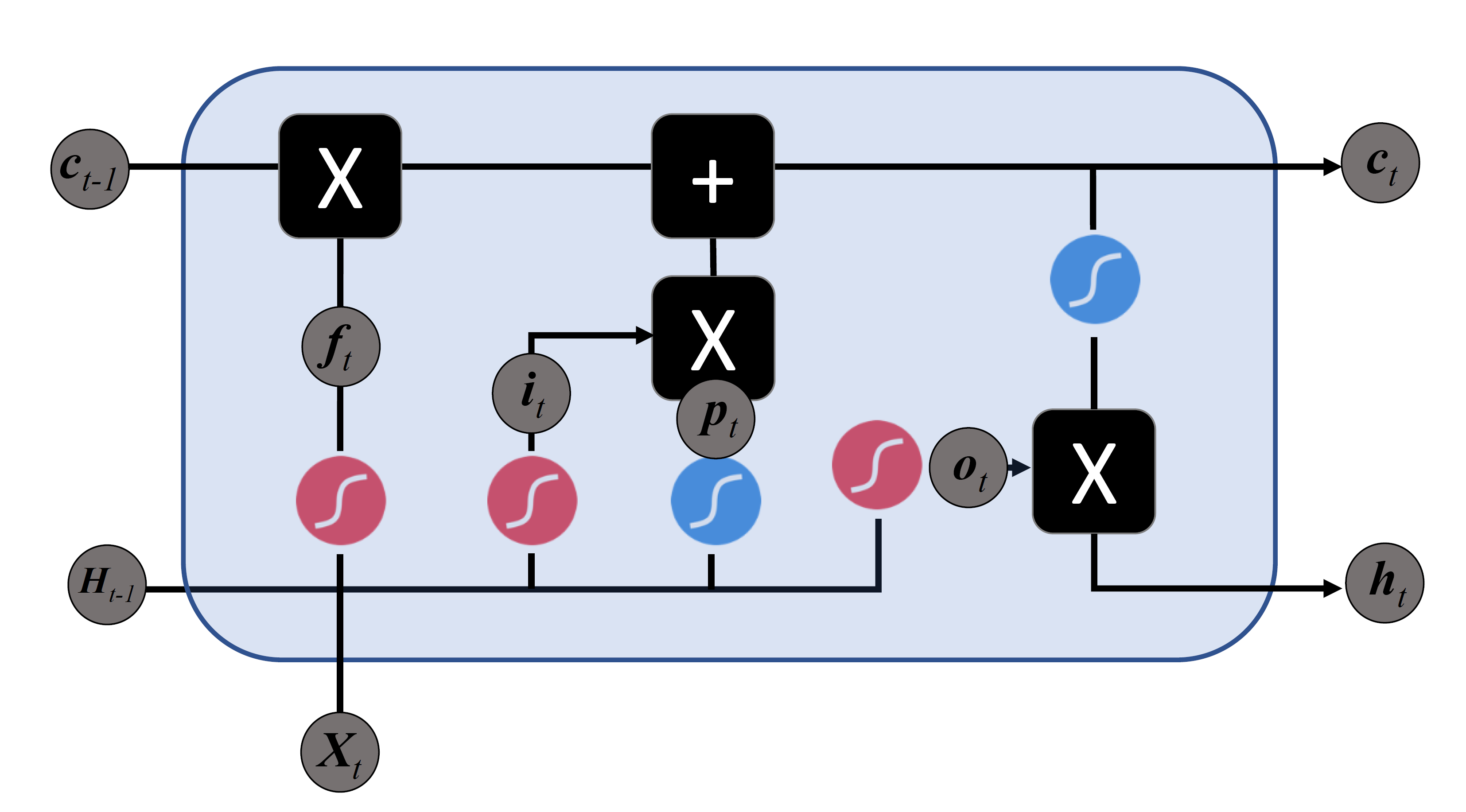}
    \caption{Information flow in a LTSM Cell}
    \label{F:LSTM2}
\end{figure}

Algorithm \ref{A:LSTM} describes how the LSTM cell works. $\b f_{t,m}$ represents the output of the forget gate. It is a combination of the previous hidden-state ($\b H_{t,m-1}$) with the new information ($\b X_{t,m}$). Note that $\b f_{t,m}\in[0,1]$ and it attenuates the signal coming com $\b c_{t,m-1}$. The input and output gates have similar structure. Their goal is to filter the ``relevant'' information from the previous minute as well as from the new input.  $\b p_{t,m}$ scales the combination of inputs and previous information. This signal will be then combined with the output of the input gate ($\b i_{t,m}$). The new hidden state will be an attenuation of the signal coming from the output gate. The prediction is a linear combination of hidden states. Figure \ref{F:LSTM2} illustrates how the information flows in a LSTM cell.

\begin{algorithm}\label{A:LSTM}
Mathematically, RNNs can be defined by the following algorithm:
\begin{enumerate}
\item
Initiate with $\b{c}_{t,0}=0$ and $\b{H}_{t,0}=0$.
\item
Given the input $\b{X}_{t,m}$, for $m\in\{1,\dots,M\}$, do:
\[
\begin{split}
\b{f}_{t,m} &= \textnormal{Logistic}(\b{W}_f \b{X}_{t,m} + \b{U}_f \b{H}_{t,m-1}+ \b{b}_f)\\
\b{i}_{t,m} &= \textnormal{Logistic}(\b{W}_i \b{X}_{t,m} + \b{U}_i \b{H}_{t,m-1}+ \b{b}_i)\\
\b{o}_{t,m} &= \textnormal{Logistic}(\b{W}_o \b{X}_{t,m} + \b{U}_o \b{H}_{t,m-1}+ \b{b}_o)\\
\b{p}_{t,m} &= \textnormal{Tanh}(\b{W}_c \b{X}_{t,m} + \b{U}_c \b{H}_{t,m-1}+ \b{b}_c)\\
\b{c}_{t,m} &= (\b{f}_{t,m} \odot \b{c}_{t,m-1}) +  (\b{i}_{t,m} \odot \b{p}_{t,m})\\
\b{h}_{t,m} &= \b{o}_{t,m} \odot \textnormal{Tanh}(\b{c}_{t,m})\\
\widehat{\b{Y}}_{t,m+h|m} &= \b{W}_y\b{h}_{t,m} +\b{b}_y
\end{split}
\]
where $\b{U}_f$, $\b{U}_i$, $\b{U}_o$ ,$\b{U}_c$ ,$\b{U}_f$, $\b{W}_f$, $\b{W}_i$, $\b{W}_o$, $\b{W}_c$, $\b{b}_f$, $\b{b}_i$, $\b{b}_o$, and $\b{b}_c$ are parameters to be estimated.
\end{enumerate}
\end{algorithm}





\subsection{Regression Trees and Random Forests}

A regression tree is a nonparametric model that approximates an unknown nonlinear function with local predictions using recursive partitioning of the space of the explanatory variables (predictors). 

The idea of regression trees is to approximate $\mathcal{F}_h(\b{X}_{t,m})$ in \eqref{eq:model} by
\[
h_D(\b{X}_{t,m}) = \sum_{j=1}^{J} \beta_j I_j(\b{X}_{t,m}),\quad\textnormal{where}\quad
I_k(\b{X}_{t,m})=
\begin{cases}
1 & \textnormal{if } \b{X}_{t,m} \in \mathcal{R}_j,\\
0 & \textnormal{otherwise}.
\end{cases}
\]
From the above expression, it becomes clear that the approximation of $\mathcal{F}_h(\cdot)$ is equivalent to a linear regression on $J$ dummy variables, where $I_j(\b{X}_{t,m})$ is a product of indicator functions.

\begin{figure}
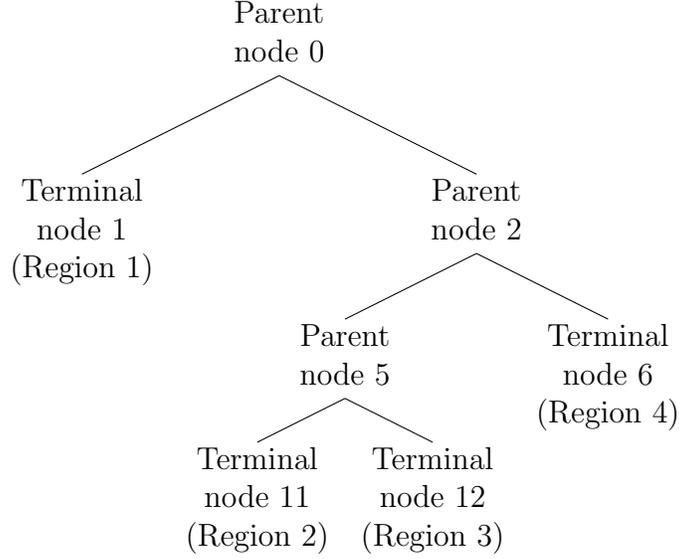

\[
\Tree [.{Parent \\ node 0} {Terminal \\ node 1\\ (Region 1)} [.{Parent \\ node 2} [.{Parent \\ node 5} {Terminal \\ node 11 \\ (Region 2)} {Terminal \\ node 12\\ (Region 3)} ] {Terminal \\ node 6\\ (Region 4)} ] ]
\]
\caption{Example of tree with labels.}
\label{F:example2}
\end{figure}

Let $J$ and $N$ be, respectively, the number of terminal nodes (regions, \emph{leaves}) and parent nodes. Different regions are denoted as $\mathcal{R}_1,\ldots,\mathcal{R}_J$. The root node at position $0$. The parent node at position $j$ has two split (child) nodes at positions $2j+1$ and $2j+2$. Each parent node has a threshold (split) variable associated, $X_{s_jt}$, where $s_j\in\mathbb{S}=\{1,2,\ldots,p\}$. Define $\mathbb{J}$ and $\mathbb{T}$ as the sets of parent and terminal nodes, respectively. Figure \ref{F:example2} gives an example. In the example, the parent nodes are $\mathbb{J}=\{0,2,5\}$ and the terminal nodes are $\mathbb{T}=\{1,6,11,12\}$.

Therefore, we can write the approximating model as
\begin{equation}
h_D(\b{X}_{t,m})=\sum_{i\in\mathbb{T}}\beta_iB_{\mathbb{J}i}\left(\b{X}_{t,m};\b{\theta}_i\right),
\end{equation}
where
\begin{equation}
B_{\mathbb{J}i}\left(\b{X}_{t,m};\b{\theta}_i\right)=
\prod_{j\in\mathbb{J}}I(X_{s_j,t,m};c_j)^{\frac{n_{i,j}(1+n_{i,j})}{2}}
\times\left[
1-I(X_{s_j,t,m};c_j)
\right]^{(1-n_{i,j})(1+n_{i,j})},
\end{equation}
\[
I(X_{s_j,t,m};c_j)=
\begin{cases}
1 & \textnormal{if}\, X_{s_j,t,m}\leq c_j\\
0 & \textnormal{otherwise},
\end{cases}
\]
\[
n_{i,j}=
\begin{cases}
-1 & \text{if the path to leaf } \,i\, \text{does not include parent node } j; \\
0  & \text{if the path to leaf } \,i\, \text{include the \textbf{right-hand} child of parent node } j; \\
1  & \text{if the path to leaf } \,i\, \text{include the \textbf{left-hand} child of parent node } j. \\
\end{cases}
\]
$\mathbb{J}_i$: indexes of parent nodes included in the path to leaf $i$. $\b{\theta}_i=\{c_k\}$ such that $k\in\mathbb{J}_i$, $i\in\mathbb{T}$ and  $\sum_{j\in\mathbb{J}}B_{\mathbb{J}i}\left(\b{X}_{t,m};\b{\theta}_j\right)=1$.

Random Forest (RF) is a collection of regression trees, each specified in a bootstrap sample of the original data. The method was originally proposed by \citet{breiman2001random}. Since we are dealing with time series, we use a block bootstrap. Suppose there are $B$ bootstrap samples. For each sample $b$, $b=1,\ldots,B$, a tree with $K_b$ regions is estimated for a randomly selected subset of the original regressors. $K_b$ is determined in order to leave a minimum number of observations in each region. The final forecast is the average of the forecasts of each tree applied to the original data:
\[
\widehat{Y}_{t,m+h|m}=\frac{1}{B}\sum_{b=1}^B\left[\sum_{i=1}^{\mathbb{T}_b} \widehat{\beta}_{i,b} B_{\mathbb{J}i,b}(\b{X}_{t,m}; \widehat{\b{\theta}}_{i,b})\right].
\]




\section{Data} \label{sec:data}

Our dataset consists of 1 minute frequency data for both the S\&P 500 ETF (SPY) and the CBOE Volatility Index (VIX), gathering a total of 3,000 business days between January 2005 and December 2016. We filter the data within each day to the observations between 09:40 and 15:50 (inclusive).

We calculate log-returns of SPY for each minute of day $t$, using a rolling five-minute return scheme, as follows:
\begin{equation}
\label{eq:logret}
    r_{t,m-4:m} = \log(P_{t,m}) - \log(P_{t,m-4}), 
\end{equation}
where $r_{t,m_2:m_1}$ represents the log difference in prices computed between minutes $m_1$ and $m_2$, and $P_{t,m}$ is the price of SPY at day $t$ and minute $m$. We explore the VIX index variable in level, and implement a transformation of the annualized VIX to an intraday minute variable, in order to facilitate the interpretation of our results\footnote{$VIX_{\text{Intraday}} = VIX_{\text{Annual}} \times (1/\sqrt{1440}) \times (1/\sqrt{252})$: 252 business days in a year and 1440 minutes in a day.}. Using this data set, we also compute the one-minute squared version VIX$_m^2$ and first difference $\Delta$VIX$_{m}$.

Inspired by \cite{BTZ2009}, we generate a measure of what could be a high-frequency analogous to the \textit{variance risk premium}, as the difference between the squared one-minute SPY return and the VIX$^2$ at minute $m$:
\begin{equation}
\label{eq:VRP}
    VRP_{t,m} = r^2_{t,m} - VIX^2_{t,m}.
\end{equation}

The following estimations will be centered on models that use volatility measures as inputs, particularly the VIX. Despite presenting additional configurations and benchmark models, we focus on understanding how the minute-by-minute VIX relates to very high-frequency market returns, as well as its overall predictive capability across the models. 

To illustrate the common behavior of market return and the volatility measure over the years, Figure \ref{fig:SPYvsVIX} displays the contemporary relation between the average minute-by-minute SPY log-return and VIX for each day of the sample. The darker the dots in the figure, the more recent the analyzed sample. It is noticeable that the relationship between both variables becomes less disperse over time, more concentrated in recent years, exhibiting lower returns for lower levels of volatility. Additionally, we present the summary statistics for minute-by-minute SPY log-returns and the annualized VIX in Table \ref{tab:summarystats}, which shows that both variables are skewed right and leptokurtic. As expected, market returns and the VIX are negatively correlated.

Despite the clear intrinsic relationship between market returns and near-term volatility implied by stock index option prices, we want to investigate the forecasting potential of SPY returns in high-frequency using volatility measures. The subsequent section presents the fundamental estimation framework adopted in this paper, based on a minute-by-minute rolling window mechanism. This approach is similar to the one developed by \cite{Chinco2019}, and allows the researcher to understand very short-term effects over intraday periodicity variables.

\section{Empirical Analysis} \label{sec:analysis}

To estimate our models, we implement a 30-minute rolling window estimation scheme for very high-frequency predictions. For each minute $m$ and day $t$, we use the previous non-overlapping thirty minute window to forecast the one-minute-ahead five-minute log-return of SPY. At the end of each day $t$, we run a total of 340 estimations between 09:40 and 15:50\footnote{We chose to avoid the first and last ten minutes of each business day in our estimations because these are the moments that usually contain a greater amount of missing values and/or repeated values.}.

It is important to point out that our high frequency analysis uses \textit{non-overlapping} input and output variables. This means that for each minute-by-minute log return $r_{t,m_2:m_1}$, we use lagged regressors that do not overlap minutes $m_2$ to $m_1$. For example, if we want to run a regression of the five-minute log return ($r_{t,m-4:m}$) on the lagged $\Delta$VIX, we compute the \textit{non-overlapping} variable as $\Delta$VIX$_{t,m-6:m-5} =$ VIX$_{t,m-5} -$ VIX$_{t,m-6}$. This strategy prevents from using future information to forecast minute $m+h$ when observations are available only up to minute $m$.

Machine learning algorithms perform better when numerical input variables are scaled. In each window, we perform a standard \textit{MinMax} scaling to the data, by using the parameters from the \textit{train} observations to scale both \textit{train} and \textit{test} data\footnote{For each window $\omega$ rescale: $\tilde{x}^{train}_{\omega,m} = \frac{x^{train}_{\omega,m} - \min(x^{train}_{\omega})}{\max(x^{train}_{\omega}) - \min(x^{train}_{\omega})}$ and  $\tilde{x}^{test}_{\omega,m} = \frac{x^{test}_{\omega,m} - \min(x^{train}_{\omega})}{\max(x^{train}_{\omega}) - \min(x^{train}_{\omega})}$}. This transformation rescales variables into the range $[0,1]$. We implement the standardization to the entire estimation data set, explored in both machine learning and benchmark models, as described in sequence. 

\subsection{Machine Learning Models}  \label{subsec:mlmodels}

The architecture of the LSTM network allows the model to learn and forecast long sequences of data, a particularly useful attribute for time-series analysis. This approach can be put into action by running rolling window regressions in a one-shot multi-step framework for each day $t$. There are many options available to the researcher in terms of network structure and hyper-parameters to be chosen, including the number of layers, hidden states, loss function and optimizer. In our paper, we optimize the hyper-parameter by tuning it according the models performance on random subsamples.

As discussed in Section \ref{subsec:nonlinearmodels}, Random Forest models (which we will also refer as RF from now on) fit classifying decision trees on data subsamples and employ an averaging approach to enhance the predictive power of the data. This ensemble learning method bootstraps the observations in blocks, using randomly sampled training sets (estimation intraday windows, in our case), controlling for potential overfit of the model. The choice of the number of trees to be used is a critical decision in a Random Forest, once it represents the main tuning engine in the model framework.

For each machine learning model (LSTM and RF), we establish \textit{three groups} of predictors $\b{X}_{t,m}$. The first one is composed exclusively by the lagged volatility variable $\b{X}_{t,m} = VIX_{t,m-5}$, while the second contains only the lagged five-minute return $\b{X}_{t,m} = r_{t,m-9:m-5}$. We call these models VIX and AR(1), respectively. Finally, the last group gathers all the variables described in Section \ref{sec:data}: 
\[
\b{X}_{t,m} = \left(r_{t,m-9:m-5},  r_{t,m-9:m-5}^2, VIX_{t,m-5}, \Delta VIX_{t,m-6:m-5}, VRP_{t,m-5}\right)',
\]
which we labeled as the aggregate model.

\subsection{Benchmark Models} \label{subsec:benchmodels}

As benchmarks, we chose different setups for $\b{X}_{t,m}$, implementing the same rolling-window approach. In addition to applying a traditional autoregressive model of order 1, we use lagged squared return, VIX, $\Delta$VIX, and $VRP$ as predictors. We estimate the coefficients in each window through standard ordinary least squares (OLS) regressions. 

We define the benchmark models as follows:

\vspace{-5mm}

\begin{align}
\label{eq:benchmarks}
\begin{split}
& \textrm{OLS-AR(1):} \quad r_{t,m-4:m} = \alpha_m + \beta_m \cdot r_{t,m-9:m-5} + \varepsilon_{t,m}
\\
& \textrm{OLS-RV:} \quad r_{t,m-4:m} = \alpha_m + \beta_m \cdot r_{t,m-9:m-5}^2 + \varepsilon_{t,m}
\\
& \textrm{OLS-VIX:} \quad r_{t,m-4:m} = \alpha_m + \beta_m \cdot VIX_{t,m-5} + \varepsilon_{t,m}
\\
& \textrm{OLS-}\Delta\textrm{VIX:} \quad r_{t,m-4:m} = \alpha_m + \beta_m \cdot \Delta VIX_{t,m-6:m-5} + \varepsilon_{t,m}
\\
& \textrm{OLS-VRP:} \quad r_{t,m-4:m} = \alpha_m + \beta_m \cdot VRP_{t,m-5} + \varepsilon_{t,m}
\end{split}
\end{align}

We use the logic presented in the beginning of this section in order to regress the market returns on \textit{non-overlapping} lagged variables. The benchmarks OLS-AR(1) and OLS-VIX are particularly important to our analysis, once it can be directly compared to the first two groups of predictors defined in the previous section. From both machine learning and benchmark models, we expect to understand the importance of the lagged log-return, as well as the VIX, in high-frequency forecasting. 

This estimation practice was executed in a cloud environment, applying
\code{scikit-learn}, \code{tensorflow}, and \code{keras} Python libraries to implement machine learning methods. We employ \code{Python 3.9.4 version} to perform the estimations and \code{R 4.1.2 version} for output analysis. In the next section, we compare the performance metrics for both machine learning and benchmark models. We contrast specifically the regressions over different groups of predictors presented in Section \ref{subsec:mlmodels} to the single input OLS models in the set of equations \ref{eq:benchmarks}. 

\section{Estimation Results} \label{sec:results}

Due to the high dimension of the estimated intraday outputs in each model (described in Section \ref{sec:analysis}), we summarise our results through two performance metrics widely used in the literature: the daily Root-mean-square error (RMSE) and \textit{out-of-sample} $R^2$. To obtain the mean and median RMSE, as well as the standard deviation, we calculate the daily measure

\begin{equation}
    RMSE_t = \sqrt{\frac{1}{M}\sum_{m=1}^{M} (r_{t,m} - \hat{r}_{t,m})^2}
\end{equation}

and then compute its overall metric throughout the sample. Then, we calculate the same \textit{out-of-sample} performance metric implemented by \cite{WelchGoyal2008} and \cite{CampbellThompson2008} for each day $t$

\vspace{-5mm}

\begin{equation}
\label{eq:oosr2eq}
    R^2_{OOS,t} =  1 - \frac{ \sum_{m=1}^{M} (r_{t,m} - \hat{r}_{t,m})^2 }{ \sum_{m=1}^{M} (r_{t,m} - \overline{r}_{t,m})^2}
\end{equation}

to generate the average $R^2_{OOS}$, its median, and standard deviation across the days. This metric is relevant to introduce a relation between the mean squared error of our model estimate and the naive projection, represented by the \textit{in-sample} historical mean within each estimation window. If the error of the model's prediction is lower than the error of the naive forecast, the $R^2_{OOS}$ is necessarily positive. The same holds for the opposite case, in which a negative $R^2_{OOS}$ represents a better performance of the historical average model. 

\subsection{Intraday Analysis} \label{subsec:intradayanalysis}

To exemplify how each predictive model works throughout a day, Figures \ref{fig:intraday_vix} and \ref{fig:intraday_ar1} illustrate the intraday rolling prediction for a randomly selected date $t$ (January 8, 2007). Each plot represents a machine learning model (LSTM and Random Forest) compared to the actual series, the historical mean based prediction, and the respective VIX or AR(1) benchmark. At each day $t$ calculate the daily performance metrics for each one of those series, and then compute the average, mean, and standard deviation for the entire sample. 

The top plot in Figure \ref{fig:intraday_vix} presents the LSTM-VIX model (that uses VIX as predictive variable) minute-by-minute predictions, while the bottom one displays the same scenario for the RF-VIX model. Comparing both to the OLS-VIX benchmark, and the naive projection, one can see that the LSTM method tracks the observed series more smoothly, but also more precisely, while the Random Forest follows pretty much the same pattern established by the benchmark model. For this specific date, the $R^2_{OOS}$ for the LSTM, Random Forest, and benchmark models were $21.15\%$, $-37.10\%$, and $6.33\%$, respectively. We can see a clear superiority of the LSTM model, followed by the benchmark and, finally, the Random Forest model.

This pattern is repeated in Figure \ref{fig:intraday_ar1}, in which we perform the same analysis comparing autoregression based models. The minute-by-minute series forecasts follow an akin behavior to that observed in the previous figure. The curves are smoother, but also more precise for LSTM-AR(1) predictions. Even more interesting, the model now presents a \textit{positive} $R^2_{OOS}$, compared to the negative ones obtained through the RF-AR(1) and OLS-AR(1) models on the same date: $6.93\%$, $-21.87\%$, and $-14.03\%$, respectively. Once again, the Random Forest output suggests a poor performance for this model. Obviously, these results are restricted to the analysis of a random day in the sample. The subsequent tables present and summarize the general outputs obtained by the estimation of the full sample, leading to similar conclusions.

\subsection{Overall performance} \label{subsec:overallperformance}

Tables \ref{tab:est_r2oos} and \ref{tab:est_rmse} present the average, median, and standard deviation of $R^2_{OOS}$ and RMSE, respectively. These outputs are calculated over the performance metrics, using \textit{out-of-sample} predictions for each day $t$, as described above\footnote{We trim the outputs in each model to restrict it to be between the 1$^{st}$ and 99$^{th}$ percentiles, excluding outliers in both directions.}. Looking only at the bottom panel of each table, which contains the benchmarks presented in the set of equations \ref{eq:benchmarks}, there is a clear hegemony of the OLS-VIX in terms of performance. The mean and median values of the \textit{out-of-sample} $R^2$ outperform the naive prediction based on the historical mean ($R^2_{OOS}>0$). As expected, this same conclusion can be drawn from the RMSE table, in which the VIX excels all the other variables as a predictor of market returns among the OLS equations.

When we include the LSTM and Random Forest machine learning models in the analysis, we can see critical improvements (and some downgrades) in terms of performance. As one can see, the LSTM-VIX is the best performing model, followed by the OLS-VIX model. Even though the RF-VIX model performs well when compared to other models, it is not able to surpass the benchmark's performance. The LSTM approach for the autoregressive component also improves its performance when compared to the OLS-AR(1) model. Differently from the benchmark outputs, the mean and median $R^2_{OOS}$ values for the LSTM-AR(1) are strictly positive, outperforming the historical mean projection. On the other hand, the RF-AR(1) model presents a drastic deterioration in performance when we use the lagged market return as predictor.  

The LSTM and RF aggregate models (including all variables) do not present any improvement in its statistics when compared to the OLS-VIX benchmark, signaling that inserting other volatility measures and lagged market returns on the same model do not provide a better performance, compared to the models presented previously. To illustrate the different distributions of $R^2_{OOS}$ metrics across the days in our sample, Figure \ref{fig:density_oosR2} presents the density for both machine learning (in the top row), and benchmark models (in the bottom). As one can see, the LSTM $R^2_{OOS}$ distribution is predominantly left-skewed for every group of predictors, in accordance with the results presented in Tables \ref{tab:est_r2oos} and \ref{tab:est_rmse}. The RF models distributions present a reasonable performance in general, apart from the RF-AR(1). Still, these models are not able to overcome the OLS-VIX, which stands out as the best performing benchmark.

In a complementary way, Figure \ref{fig:MLvsBench_RMSE} pursues to unravel the relationship between the results obtained by each machine learning model, compared to their respective benchmarks over the years. The vertical axis of each plot represent the LSTM and Random Forest models estimations using as regressor the AR(1) (in the first column of plots) and the VIX (in the second column). The horizontal axis displays the predicted values for the benchmark models. This visualization allows one to identify when machine learning models were able to overcome its benchmarks (the points below the \ang{45} dotted line). The last years in our sample seem to present lower and less disperse estimation errors for all the models combinations. Moreover, the VIX-based models got even better recently, when using both LSTM and RF models. These highlights synthesize the objective of this paper: contribute to the short-term forecasting literature, as well as the implementation of modern forecasting techniques through nonlinear rolling window models in a high-frequency setting.

\section{Conclusion} \label{sec:conclusion}

There is growing interest in understanding how to predict high-frequency stock returns. Our paper introduces an innovative approach to estimate minute-by-minute market return (SPY) forecasts using volatility measures and lagged returns as predictors. In addition to standard benchmark models, we implement nonlinear machine learning methods as an attempt to capture idiosyncrasies of this kind of data, whose properties can differ substantially from variables at lower frequencies. 

The outputs obtained in our estimations indicate a preliminary, but also encouraging path to better understand how to predict high-frequency market returns. We focus on neural networks (Long-Short-Term Memory) and tree-based (Random Forests) models to estimate multiple intraday rolling windows, using different regressors configurations. Models that use the Cboe Volatility Index (VIX) as predictor stand out, indicating that the VIX is a strong candidate predictor, when compared to other variables. The precision of the forecasts obtained using the OLS benchmark model gets even higher when we apply the LSTM to predict market returns using exclusively the VIX. Although the random forest model is not able to outperform the results obtained by the previous models (eventually even worsening it in some cases), the LSTM proved to be very promising in terms of predictive power, for both VIX and past market returns as regressors.

\clearpage

\bibliographystyle{chicago}
\bibliography{ref}

\clearpage

\begin{appendices}
\counterwithin{figure}{section}
\counterwithin{table}{section}

\section{Appendix: Figures}
\label{appendix:figures}

\vfill

\begin{figure}[!htbp]
    \caption{Average $r_{t,m}$ and $VIX_{t,m}$ at each day $t$}
    \label{fig:SPYvsVIX}
    \centering
    \includegraphics[scale=.35]{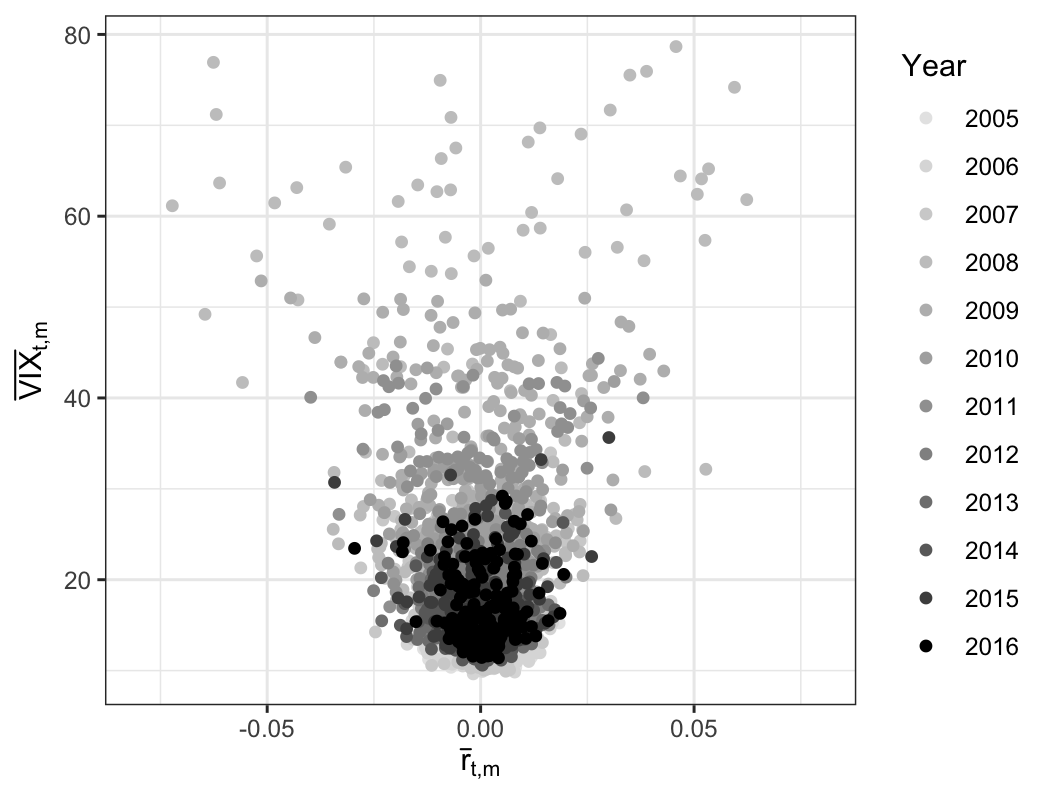}
    
    \begin{minipage}{.75\textwidth} 
    
    {\footnotesize The figure presents the daily average $r_{t,m}$ and $VIX_{t,m}$ computed minute-by-minute within each day. The color scale represents the different years analyzed in the sample. The sample period extends from January 2005 to December 2016.}
    
    \end{minipage}
    
\end{figure}

\vfill

\begin{figure}[!htbp]
    \caption{Intraday prediction: VIX models (January 8, 2007)}
    \label{fig:intraday_vix}
    \centering
    \includegraphics[scale=.42, trim=3cm 3cm 0cm 4cm]{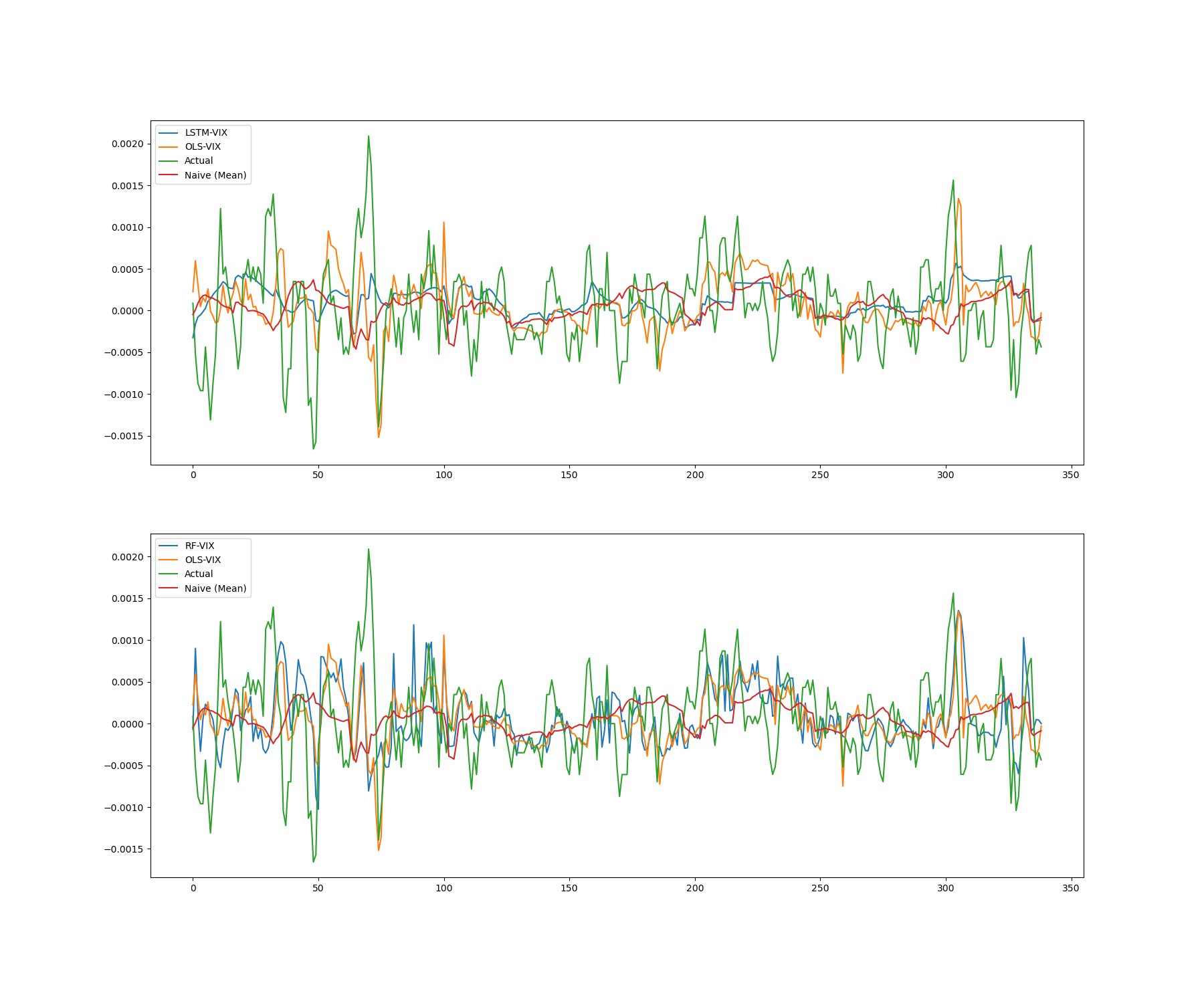}
    
    \begin{minipage}{.95\textwidth} 
    
    {\footnotesize The plots above represent the intraday rolling prediction of minute-by-minute log-returns on the market. As detailed in Section \ref{sec:analysis}, we run 340 estimations between 09:40 and 15:50, where the first prediction takes place at 10:11 (after the first 30 minute window). The plots represent the LSTM and Random Forest VIX-based models compared to its respective OLS-VIX benchmark, as well as the naive projection. The sample period extends from January 2005 to December 2016.}
    
    \end{minipage}
    
\end{figure}

\vfill

\begin{figure}[!htbp]
    \caption{Intraday prediction: AR(1) models (January 8, 2007)}
    \label{fig:intraday_ar1}
    \centering
    \includegraphics[scale=.42, trim=3cm 3cm 0cm 4cm]{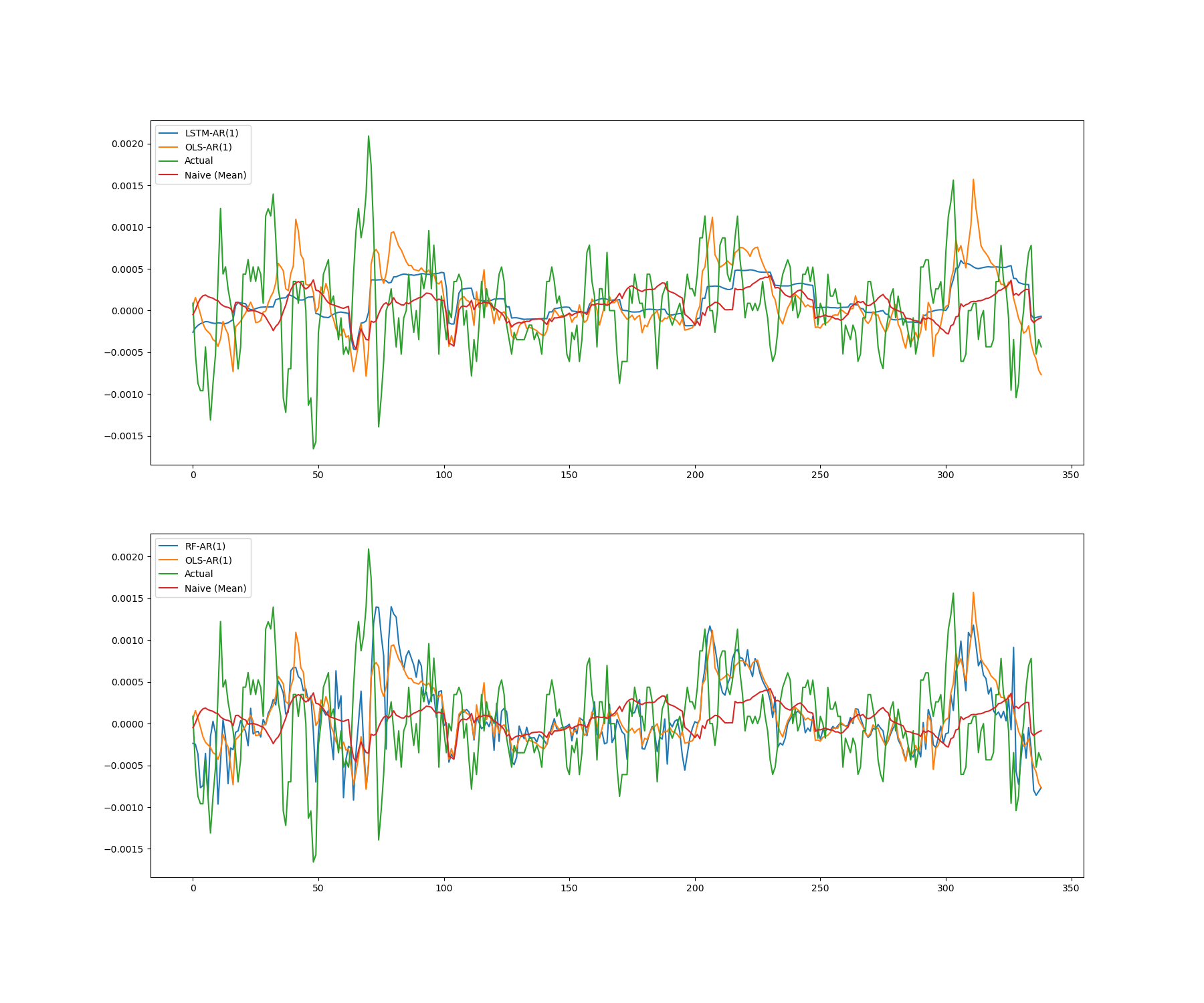}
    
    \begin{minipage}{.9\textwidth} 
    
    {\footnotesize The plots above represent the intraday rolling prediction of minute-by-minute log-returns on the market. As detailed in Section \ref{sec:analysis}, we run 340 estimations between 09:40 and 15:50, where the first prediction takes place at 10:11 (after the first 30 minute window). The plots represent the LSTM and Random Forest AR(1)-based models compared to its respective OLS-AR(1) benchmark, as well as the naive projection. The sample period extends from January 2005 to December 2016.}
    
    \end{minipage}
    
\end{figure}

\vfill

\begin{figure}[!htbp]
    \caption{$R^2_{OOS}$ distribution and density plots}
    \label{fig:density_oosR2}
    \centering
    \includegraphics[scale=.17]{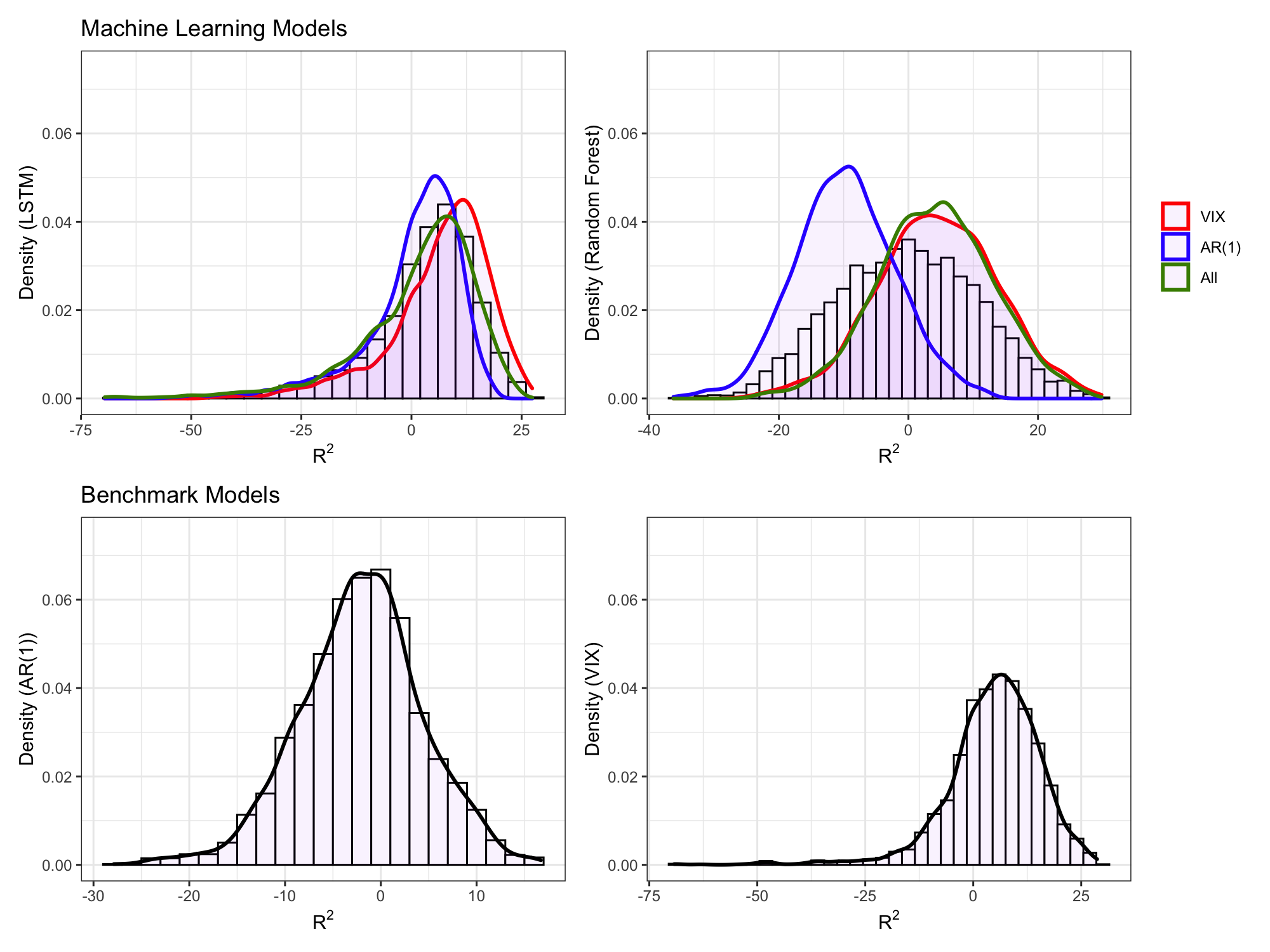}
    
    \begin{minipage}{.9\textwidth} 
    
    {\footnotesize The figure presents the density plots for the daily $R^2_{OOS}$ distribution generated by each model. The first row of plots display machine learning models, divided by the groups of regressors described in Section \ref{subsec:mlmodels}, while the second row exhibits benchmark OLS models detailed in equations \ref{eq:benchmarks}. The sample period extends from January 2005 to December 2016.}
    
    \end{minipage}
    
\end{figure}

\vspace{-5mm}

\begin{figure}[!htbp]
    \caption{Daily RMSE: Machine Learning models \textit{vs} Benchmarks}
    \label{fig:MLvsBench_RMSE}
    \centering
    \includegraphics[scale=.18]{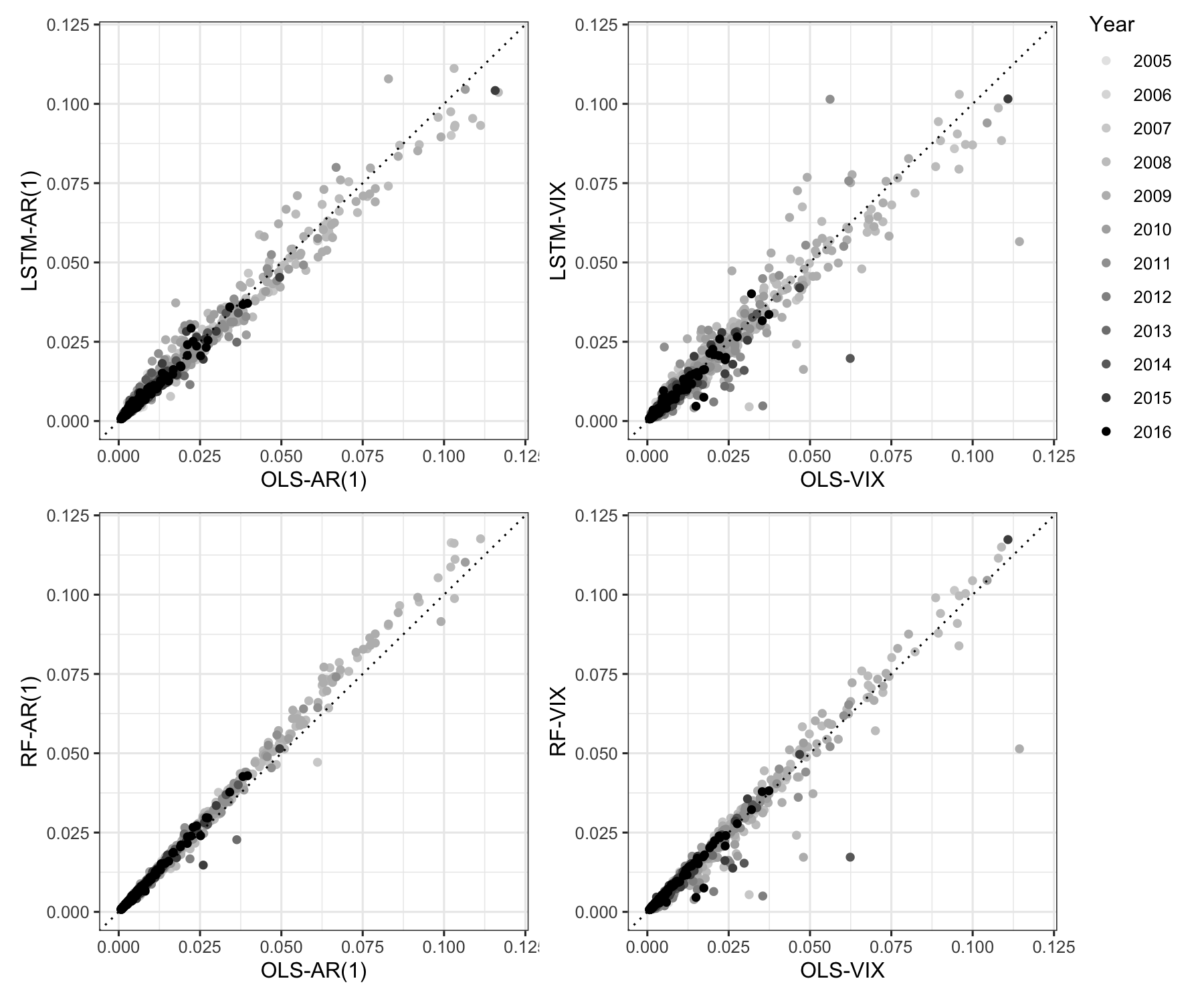}
    
    \begin{minipage}{.9\textwidth} 
    
    {\footnotesize The plots above represent each machine learning model (vertical axis) against its respective benchmark (horizontal axis). The first column of plots displays the OLS-AR(1) model, while the second reports the OLS-VIX model. The rows of plots represent the LSTM and Random Forest models using AR(1) and VIX as single regressors, respectively. The sample period extends from January 2005 to December 2016.}
    
    \end{minipage}
    
\end{figure}

\clearpage

\section{Appendix: Tables}
\label{appendix:tables}

\vspace{5cm}

\begin{table}[!htbp] \centering 
  \caption{Summary Statistics} 
  \label{tab:summarystats} 
  
\begin{threeparttable}
\centering 
\begin{tabular}{@{\extracolsep{5pt}} lcc} 
\\[-1.8ex]\hline 
\hline \\[-1.8ex] 
Variable & SPY & VIX \\ 
\hline \\[-1.8ex] 
\textit{Descriptive Statistics} (\%) & & \\
\hline
Mean & $0.0001$ & $19.519$ \\ 
Std. Deviation & $0.099$ & $9.404$ \\ 
Skewness & $0.168$ & $2.483$ \\ 
Kurtosis & $42.886$ & $8.140$ \\ 
\hline
\textit{Correlations} (\%) & & \\
\hline
SPY & $100$ & $-0.432$ \\ 
VIX & $-0.432$ & $100$ \\  
\hline \\[-1.8ex] 
\end{tabular} 

\footnotesize

The summary statistics and correlations are reported for intraday variables in percentage form for both the minute by minute SPDR S\&P 500 ETF Trust (SPY) and the annualized Cboe Volatility Index (VIX) at minute frequency. The sample period extends from January 2005 to December 2016.

\end{threeparttable}

\end{table} 

\clearpage

\begin{table}[!htbp] \centering 
  \caption{Out-of-sample $R^2$} 
  \label{tab:est_r2oos} 
  
\resizebox{\textwidth}{!}{  
\begin{tabular}{@{\extracolsep{5pt}} lccccccccc}
\\[-1.8ex]\hline 
\hline \\[-1.8ex] 
$R^2_{OOS} \quad (\%)$ & \multicolumn{3}{c}{Mean} & \multicolumn{3}{c}{Median} & \multicolumn{3}{c}{Std. Deviation} \\ \cmidrule{2-4} \cmidrule{5-7} \cmidrule{8-10} 

& VIX & AR(1) & Agg. & VIX & AR(1) & Agg. & VIX & AR(1) & Agg. \\ 
\hline
LSTM & 6.52 & 0.41 & 1.68 & 8.76 & 2.89 & 4.51 & 11.46 & 11.25 & 13.45 \\ 
RF & 4.45 & -9.73 & 4.17 & 4.43 & -9.75 & 4.32 & 9.27 & 7.77 & 8.86 \\ 
\hline
OLS-AR(1) & - & -2.11 & -2.11 & - & -1.99 & -1.99 & - & 6.46 & 6.46 \\ 
OLS-RV & - & - & -14.07 & - & - & -7.20 & - & - & 28.22 \\ 
OLS-VIX & 4.76 & - & 4.76 & 5.80 & - & 5.80 & 10.81 & - & 10.81 \\ 
OLS-$\Delta$VIX & - & - & -3.42 & - & - & -2.98 & - & - & 7.52 \\ 
OLS-VRP & - & - & -14.11 & - & - & -7.20 & - & - & 28.44 \\ 
\hline \\[-1.8ex] 
\end{tabular} }

\justifying
\footnotesize
\noindent The table reports mean, median, and standard deviation of the \textit{out-of-sample} $R^2$ ($R^2_{OOS}$) throughout the sample. The top panel reports the estimation outputs of machine learning models, while the bottom panel displays the ordinary least squares regression benchmark models, detailed in Section \ref{sec:analysis}. For each metric, the columns represent the groups of predictors described in Section \ref{subsec:mlmodels}: VIX, AR(1), and Aggregate. The sample period extends from January 2005 to December 2016.

\end{table} 

\vspace{3cm}

\begin{table}[!htbp] \centering 
  \caption{Root-mean-square error (RMSE)} 
  \label{tab:est_rmse} 
  
\resizebox{\textwidth}{!}{  
\begin{tabular}{@{\extracolsep{5pt}} lccccccccc} 
\\[-1.8ex]\hline 
\hline \\[-1.8ex] 
$RMSE \quad (\times 10^4)$ & \multicolumn{3}{c}{Mean} & \multicolumn{3}{c}{Median} & \multicolumn{3}{c}{Std. Deviation} \\ \cmidrule{2-4} \cmidrule{5-7} \cmidrule{8-10} 

& VIX & AR(1) & Agg. & VIX & AR(1) & Agg. & VIX & AR(1) & Agg. \\ 
\hline
Naive & 0.0085 & 0.0086 & 0.0086 & 0.0043 & 0.0043 & 0.0043 & 0.0125 & 0.0129 & 0.0127 \\ 
LSTM & 0.0080 & 0.0086 & 0.0084 & 0.0041 & 0.0044 & 0.0043 & 0.0117 & 0.0127 & 0.0123 \\ 
RF & 0.0081 & 0.0095 & 0.0082 & 0.0041 & 0.0048 & 0.0041 & 0.0121 & 0.0143 & 0.0123 \\ 
\hline
OLS-AR(1) & - & 0.0087 & 0.0087 & - & 0.0044 & 0.0044 & - & 0.0128 & 0.0128 \\ 
OLS-RV & - & - & 0.0098 & - & - & 0.0051 & - & - & 0.0145 \\ 
OLS-VIX & 0.0081 & - & 0.0081 & 0.0041 & - & 0.0041 & 0.0121 & - & 0.0121 \\ 
OLS-$\Delta$VIX & - & - & 0.0089 & - & - & 0.0046 & - & - & 0.0131 \\ 
OLS-VRP & - & - & 0.0098 & - & - & 0.0051 & - & - & 0.0145 \\ 
\hline \\[-1.8ex] 
\end{tabular} }

\justifying
\footnotesize
\noindent The table reports mean, median, and standard deviation of the Root-mean-square error (RMSE) throughout the sample. The top panel reports the estimation outputs of machine learning models, while the bottom panel displays the ordinary least squares regression benchmark models, detailed in Section \ref{sec:analysis}. For each metric, the columns represent the groups of predictors described in Section \ref{subsec:mlmodels}: VIX, AR(1), and Aggregate. The sample period extends from January 2005 to December 2016.

\end{table} 

\end{appendices}

\end{document}